\documentclass{aa}  

\usepackage{float}
\usepackage{graphicx}
\usepackage{txfonts}
\usepackage{mathptmx}
\usepackage{hyperref}
\hypersetup{colorlinks,
citecolor=blue
}
\newcommand{\ttt}[1]{{\texttt{#1}}}
\renewcommand{\d}{\mathrm{d}}

\newcommand{\kms}{\mbox{km\,s$^{-1}$}}
\newlength{\imagew}
\newlength{\imageh}
\newlength{\legendw}
\newlength{\legendh}
\newlength{\legendx}
\newlength{\legendy}
\newcommand{\graphicswithlegend}[6]{
  \setlength{\imagew}{#1}
  \settoheight{\imageh}{\includegraphics[width=\imagew]{#2}}

  \setlength{\legendw}{#3\imagew}
  \settoheight{\legendh}{\includegraphics[width=\legendw]{#4}}

  \setlength{\legendx}{\imagew}
  \addtolength{\legendx}{-\legendw}
  \addtolength{\legendx}{-#5\imagew}

  \setlength{\legendy}{\imageh}
  \addtolength{\legendy}{-\legendh}
  \addtolength{\legendy}{-#6\imageh}

  \includegraphics[width=\imagew]{#2}%
  \llap{
    \hspace{-\the\legendx}
    \raisebox{\legendy}{\includegraphics[width=\legendw]{#4}}
    \hspace{\the\legendx}
  }
}

\begin{document}

\title{Challenges in modeling the dark matter halo of NGC 1052-DF2: Cored versus cuspy halo models}

\author{K. Aditya\inst{1}}
\institute{Indian Institute of Astrophysics, Koramangala, Bengaluru 560 034, INDIA\\
\email{aditya.k@iiap.res.in}}

 
\abstract
{}
{The discovery of NGC 1052-DF2 and subsequent modeling have shown that NGC 1052-DF2 is deficient in dark matter 
and is in conflict with the standard stellar-to-halo mass ratio. In this work, we aim to resolve the degeneracy 
between the dynamical models on the mass estimate of the NGC 1052-DF2.}
{We constructed mass models of NGC 1052-DF2 using an anisotropic 
distribution function with a radially varying anisotropy parameter 
and studied the effect of the various model parameters on the dark 
matter estimates. {We used the observed stellar photometry as an input parameter to construct 
the distribution function and employed a Markov Chain Monte Carlo (MCMC) method to estimate the dark matter model parameters.}}
{We find that mass models with a cuspy dark matter halo have comparable $\chi^{2}$ to models with zero dark matter. Moreover, the cuspy dark matter halo fails to consistently account for the observed velocity dispersion in the inner and outer regions of the galaxy. Consequently, we rule out the possibility of a cuspy dark matter halo for describing the mass models of NGC 1052 - DF2. Our study shows that the cored dark matter halo model with a total mass of
$\log(M_{DM}/M_{\odot})=10.5$ explains the observed kinematics but requires an extraordinarily large scale length (20kpc) 
and an outer cutoff radius (26kpc). {While the cored mass model provides a comparatively better fit, 
our findings emphasize that the mass models are largely unconstrained by the available kinematic data. } 
{Our results suggest that NGC 1052 - DF2 may not only have an ultra-diffuse stellar distribution but that it can, within uncertainties in the available kinematic data, potentially host an ultra-diffuse dark matter distribution compatible with the standard stellar-to-halo mass relation (SHMR) 
predicted by galaxy formation and evolution models }}
{}


\maketitle
%
\section{Introduction}
Ultra-diffuse galaxies (UDGs) are a subclass of low-surface-brightness galaxies
with the characteristic luminosity of a dwarf galaxy but an effective radius larger than that 
expected for their mass content, and hence ultra-diffuse \citep{van2015forty}. 
Ultra-diffuse galaxies are defined as having g-band central surface brightnesses, 
$\mu_{g}(0)> 24\,mag\,arcsec^{-2}$, and effective radii, $R_{e}>1.5\,kpc$. 
Deep photometric observations have identified UDGs in varied environments, 
from groups to clusters and fields \citep{koda2015approximately,yagi2016catalog, 
van2016abundance, 2017MNRAS.468..703R,trujillo2019distance,2019MNRAS.485.1036M, 
2020ApJ...894...75L,2021A&A...654A.105M}. \\
In this work, we study the detailed mass models of NGC 1052 - DF2. {The dark 
matter content in this galaxy has been claimed to be in tension with the standard 
stellar-to-halo mass relation (SHMR) \citep{behroozi2010comprehensive} 
but is instead consistent with zero dark matter content.}
In this work, we show that the mass estimates are critically dependent on 
the shape and structure of the dark matter halo and anisotropy of the stellar dispersion.
With a large number of UDGs being discovered in deep 
surveys \citep{koda2015approximately,mihos2015galaxies,merritt2016dragonfly,
2021A&A...654A.105M,zaritsky2022systematically}, the detailed mass models 
presented in this work help us disentangle the role of different structural and 
dynamical parameters on the mass estimates of the dark matter halos in these systems {\citep{2021ApJ...909...20S,2022MNRAS.514.3329M,mancera2024exploring}}.

NGC 1052-DF2 was previously identified by \citet{fosbury1978active} and 
\citet{karachentsev2000dwarf} and more recently by \citet{van2018galaxy, 
emsellem2019ultra, fensch2019ultra, danieli2019still} 
as a satellite galaxy of the elliptical galaxy NGC 1052. 
The dynamical mass of the galaxy within a given radius was estimated with 
the mass estimator method (MTE) \citep{watkins2010masses} using the observed 
kinematics of ten globular clusters of the galaxy. The MTE predicts 
the total dynamical mass within 7.6 kpc to be less than 
$\rm 3.4\times 10^{8}M_{\odot}$, which is on the order of 
the estimated stellar mass of $\rm 2 \times 10^{8} M_{\odot}$. {This 
surprising result implies that within 7.6kpc the mass of the galaxy
is mostly baryonic matter, and that the dark matter needed to 
explain hierarchical structure formation in the  
$\Lambda$CDM  scenario is significantly less than expected in NGC 1052 - DF2.}\\

The $\Lambda$CDM theory gives a remarkable description of the large-scale structure of galaxies \citep{springel2006large} 
but shows discrepancies between simulations and observations \citep{bullock2017small} at galactic and sub-galactic scales. 
Using the zoom-in cosmological hydrodynamical simulations, it has been shown that UDGs like NGC 1052-DF2  
are derivatives of dwarf galaxies that have had their gas removed and star formation quenched \citep{haslbauer2019galaxies, 
tremmel2020formation, liao2019ultra, chan2018origin}. Further, the lack of dark matter in NGC 1052-DF2 has opened up avenues for testing 
alternative or modified gravity theories, for example (see \cite{moffat2019ngc, islam2020enigmatic, muller2019predicted, famaey2018mond}). Thus, 
NGC 1052-DF2 poses interesting questions regarding the formation of galaxies with minimal dark matter content, astrophysical 
processes regulating the formation of such galaxies, and the potential nature of dark matter.

Previous studies on dynamical modeling of NGC 1052-DF2 by \cite{wasserman2018deficit,hayashi2018effects} rely on 
solutions of spherically symmetric Jeans equations with a constant anisotropy parameter. Similarly, \cite{nusser2019towards} 
adopt a distribution function-based approach to estimating the dark matter mass consistent with the observed line of 
sight stellar dispersion. \cite{wasserman2018deficit} find that the models assuming the standard SHMR predict a 
large central velocity dispersion, whereas the models with halo mass kept as a free parameter describe the observation better. 
Studies by \cite{wasserman2018deficit} and \cite{hayashi2018effects} find a similar mass estimate for 
the dark matter halo $(3.4\times 10^{8}M_{\odot})$.  The large number of parameters involved in dynamical
modeling using spherical Jeans equations is riddled with degeneracies. Although detailed modeling methods exist 
for disentangling the degeneracy between different parameters \citep{mamon2010kinematic,richardson2014dark,read2017break}, 
these methods require information about the kinematics of multiple stellar populations and higher-velocity moments. 
In this study, we construct dynamical models of NGC 1052-DF2 using a distribution function-based approach 
with a radially varying anisotropy parameter \citep{cuddeford1991analytic,vasiliev2019agama}. 

The paper is organized as follows: in Sect. 2, we present the modeling methods using the self-consistent method implemented in  
 \href{https://github.com/GalacticDynamics-Oxford/Agama}{{\ttt{AGAMA}}}\footnote{https://github.com/GalacticDynamics-Oxford/Agama}  
 \citep{vasiliev2019agama}; in Sect. 3, we present the results; and finally we discuss the results and conclude in Sect. 4.

\section{Dynamical model of NGC 1052-DF2}
{NGC 1052-DF2, a UDG, has been the subject of intense study due to its unusual dark matter content. The dark matter in 
NGC 1052-DF2  appears to be significantly lower than what is predicted by the standard SHMR. 
This study aims to resolve the discrepancies in the mass modeling of NGC 1052-DF2 by employing dynamical models 
based on an anisotropic distribution function. We constructed distribution function-based models using the stellar density as an input parameter and constrained the 
parameters corresponding to the dark matter halo using the Markov Chain Monte Carlo (MCMC) method. We used the observed velocity 
dispersion profile derived by \citep{wasserman2018deficit,van2018galaxy} as a constraint on our model. We now describe the input parameter needed to construct the mass models of NGC 1052 - DF2.} 

\subsection{Stellar density}
The structure of the stellar body in NGC 1052-DF2 was parameterized by a {S\'{e}rsic} profile {\citep{sersic1968atlas}}:
\begin{equation}
\Sigma = \Sigma_0 \exp\big[-b_n\,(R/R_e)^{1/n}\big]
,\end{equation}
with a {S\'{e}rsic} index of $\rm n=0.6$, axes ratio of $\rm b/a=0.85$, and effective radius of $\rm R_{e}=2.2 kpc$ \citep{van2018galaxy}. 
The stellar mass of NGC 1052-DF2 is $\rm M_{stars}= 2.2\times10^{8} M_{\odot}$, {assuming a mass-to-light ratio 
equal to 2.0 in the V band}. The stellar density profile constitutes an input to the mass models of NGC 1052- DF2 and was 
implemented as a \ttt{Sersic} model in \ttt{AGAMA}.

\subsection{Stellar velocity dispersion}
We used the stellar dispersion profile of the NGC 1052-DF2 given in 
\cite{van2018galaxy, wasserman2018deficit}. Using the Keck Cosmic Web Imager (KCWI), \cite{danieli2019still} 
report stellar dispersion to be equal to $\sigma=8.5^{+2.3}_{-3.1}$\kms, whereas in another independent study, 
\cite{emsellem2019ultra} report stellar dispersion equal to $\sigma=10.8^{+3.2}_{-4.0}$ \kms using the Multi-Unit 
Spectroscopic Explorer (MUSE) at the ESO Very Large Telescope (VLT). {We used the observed stellar dispersion profile 
derived by \cite{wasserman2018deficit,van2018galaxy} as a constraint on the dynamical model of NGC 1052 - DF2.}

\subsection{Dark matter density }
We parameterized the dark matter profile using the general Hernquist model, which can mimic both the cored dark matter density and the cuspy one \citep{hernquist1990analytical, zhao1996analytical}:
\begin{equation}
\rho = \rho_0  \left(\frac{R}{R_{s}}\right)^{-\gamma} \Big[ 1 + \big(\frac{R}{R_{s}}\big)^\alpha \Big]^{\frac{\gamma-\beta}{\alpha}}\times \exp\Big[ -\big(\frac{R}{R_\mathrm{cut}}\big)\Big] 
,\end{equation}
where $\rho_{0}$ and $R_{s}$ are the central density and scale length, respectively, $\alpha$ is the sharpness parameter of the transition from the inner slope, $\gamma$, to the outer slope, $\beta$, and $R_{cut}$ is the outer cutoff radius. The density profile was implemented as a \ttt{Spheroid} model in  \ttt{AGAMA}. The \ttt{Spheroid} model in \ttt{AGAMA} can also be initialized with the total dark matter
mass $(M_{DM})$ within $5.3R_{s}$ using the argument \ttt{mass}. 
{In the case of the cuspy profile, the total dark matter mass $(M_{DM})$ is related to the virial mass $(M_{vir})$ and the concentration parameter $(c_{vir})$ by $M_{virial}= M_{DM}[\ln(1+c_{vir}) - c_{vir}/(1+c_{vir})]$, where $c_{vir}$ is related to the virial radius $(R_{vir})$ and the scale length $(R_{vir})$ of the dark matter halo through  $c_{vir}=R_{vir}/R_{s}$. However, in the case of cored halo, $M_{DM}$ is mass enclosed inside $5.3R_{s}$.} We modeled the dark matter density as either a  cored density profile ($\alpha=2, \,\beta=2, \, \gamma=0$) or a cuspy one ($\alpha=1, \, \beta=3, \, \gamma=1$) in our model.

\subsection{Galaxy model} 
 We combined the density of the stellar component and dark matter models to compute the total potential of the system using the 
\ttt{AGAMA} class \ttt{Potential}. The \ttt{Potential} class in \ttt{AGAMA} uses multipole expansion to compute the potential for any given density profile.
The density was decomposed into spherical harmonics with radially varying amplitudes, and finally the potential corresponding to each spherical 
harmonics term was combined to compute the total potential (for details, see \cite{vasiliev2019agama}). 
Once the potential-density pairs were known, we inverted the density profiles to obtain the distribution function with the following form \citep{cuddeford1991analytic}:
\begin{equation}  
f(E,L) = \hat f(Q) \; L^{-2\beta_0}, \qquad Q = E + L^2 / (2 R_a^2) .
\end{equation}
The above distribution function produces a velocity anisotropy, 

\begin{align}
\beta(r) \equiv 1 - \frac{\sigma_t^2}{2\sigma_r^2} = \frac{\beta_0 + (R/R_a)^2}{1 + (R/R_a)^2},
\end{align}

\noindent where $\sigma_t^2$and $\sigma_r^2$ are the tangential and radial dispersions.
In the above equation, $\beta_0$ is the anisotropy value at the center and $R_{a}$ is the anisotropy radius. The isotropic 
case was obtained by setting $\beta_0=0, \, R_a=\infty$. $\hat f(Q)$ was obtained through a more general Eddington inversion formula derived by \cite{osipkov1979spherical,merritt1985spherical,cuddeford1991analytic,vasiliev2019agama}:
\begin{equation}
\begin{aligned}
\hat f(Q) &= \left\{  \begin{array}{l}\displaystyle  \frac{2^{\beta_0}}{(2\pi)^{3/2}\, \Gamma(1-\beta_0)\; \Gamma(3/2-\beta_{0})}
\int_Q^0 \frac{\d \hat\rho}{\d \Phi} \frac{\d \Phi}{(\Phi - Q)^{3/2-\beta_0}}, \cr
 \cr
 [1/2<\beta_0<1] ,\\
\cr
\displaystyle \frac{1}{2\pi^2}\, \left. \frac{\d \hat\rho}{\d \Phi} \right|_{\Phi=Q}, [\beta_0=1/2],\\
\cr
\displaystyle \frac{2^{\beta_0}}{(2\pi)^{3/2}\, \Gamma(1-\beta_0)\; \Gamma(1/2-\beta_{0} )}\,
\int_Q^0 \frac{\d ^2\hat\rho}{\d \Phi^2} \frac{\d \Phi}{(\Phi - Q)^{1/2-\beta_0}}, \\
\cr
[-1/2<\beta_0<1/2],\\
\cr
\displaystyle \frac{1}{2\pi^2}\, \left. \frac{\d ^2\hat\rho}{\d \Phi^2} \right|_{\Phi=Q} , [\beta_0=-1/2].
\end{array} \right.
\end{aligned}
\end{equation}In the above equations, $\hat\rho$ is called augmented density and is expressed as a function of potential,
\begin{equation}
\hat\rho(\Phi) \equiv \left. \rho(r) \,r^{2\beta_0}\, \big[1 + (r/r_a)^2\big]^{1-\beta_0} \right|_{r=r(\Phi)}.    
\end{equation}

Once the total density of the system was self-consistently defined, we used the \ttt{QuasiSpherical} model implemented in the
\ttt{DistributionFunction} class in \ttt{AGAMA} to numerically invert the potential in terms of the density and numerically estimate 
the distribution function. In the above equation, $\Phi$ is the total potential obtained by adding the potential of the stellar $(\Phi_{s})$ and 
the dark matter $(\Phi_{DM})$ component.

Finally, the line-of-sight velocity dispersion was computed 
using the \ttt{moments} module implemented in \ttt{GalaxyModel} class. The numerical details about the implementation can be found 
in detail in \cite{vasiliev2018agama,vasiliev2019agama}. We used the observed line-of-sight velocity dispersion as a constraint on 
the model and compared it with the dispersion profile computed using \ttt{AGAMA}. We compared the model dispersion to the data 
using $\chi^{2}$, defined as
\begin{equation}
 \chi^{2} =\sum _{R} \frac{\bigg(\sigma_{\rm{obs}}(R) - \sigma_{\rm{\ttt{AGAMA} }}(R) \bigg)^{2} }{s^{2}_{\rm{err}}(R)},
\end{equation}
where $\sigma_{\rm{obs}}$ is the observed stellar dispersion, $\sigma_{\rm{\ttt{AGAMA} }}$ is the 
modeled line-of-sight dispersion, and $s^{2}_{err}$ is  the error on the observed dispersion.

\section{Results}

\begin{figure}
\hspace{-3mm}
\graphicswithlegend{80mm}{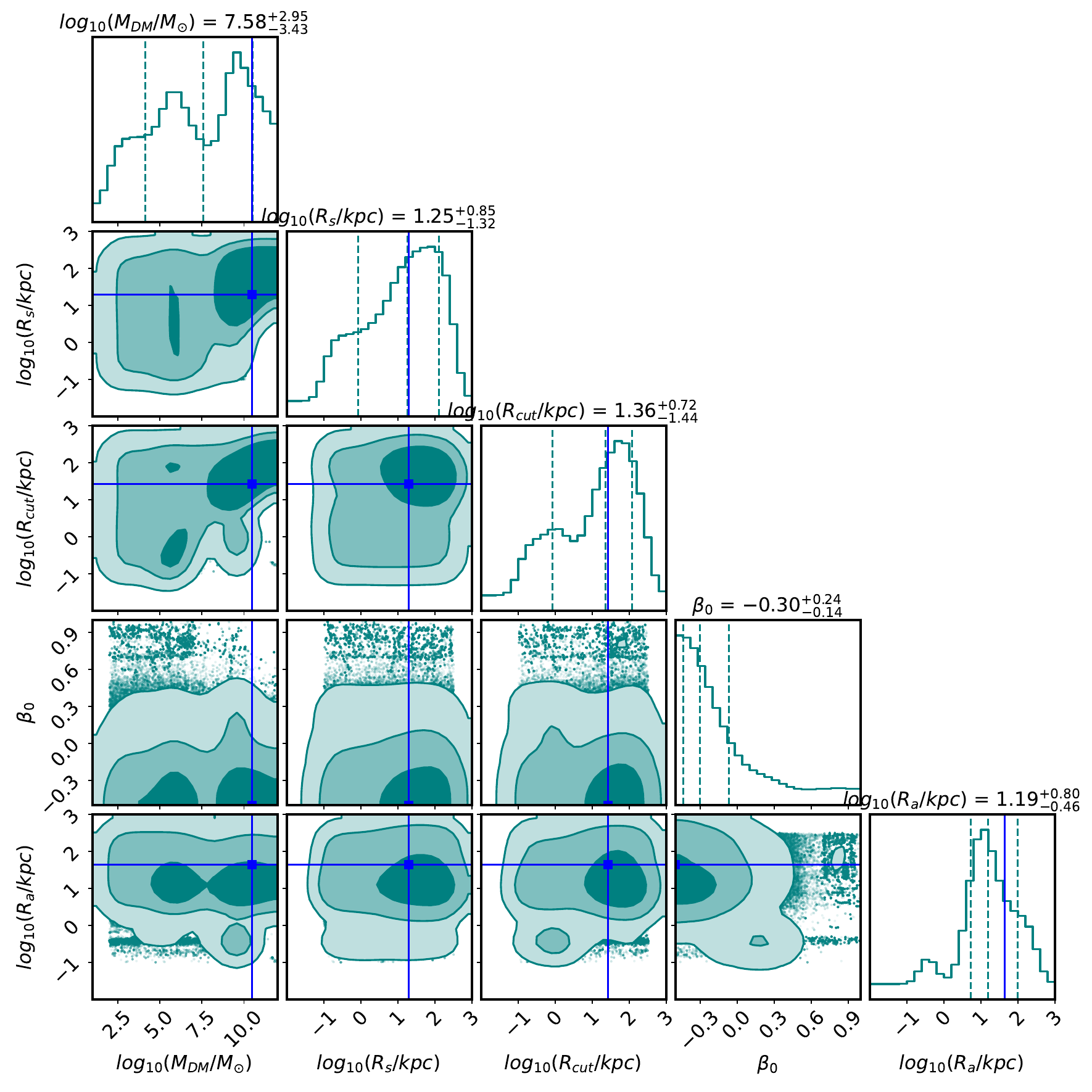}
                       {0.55}{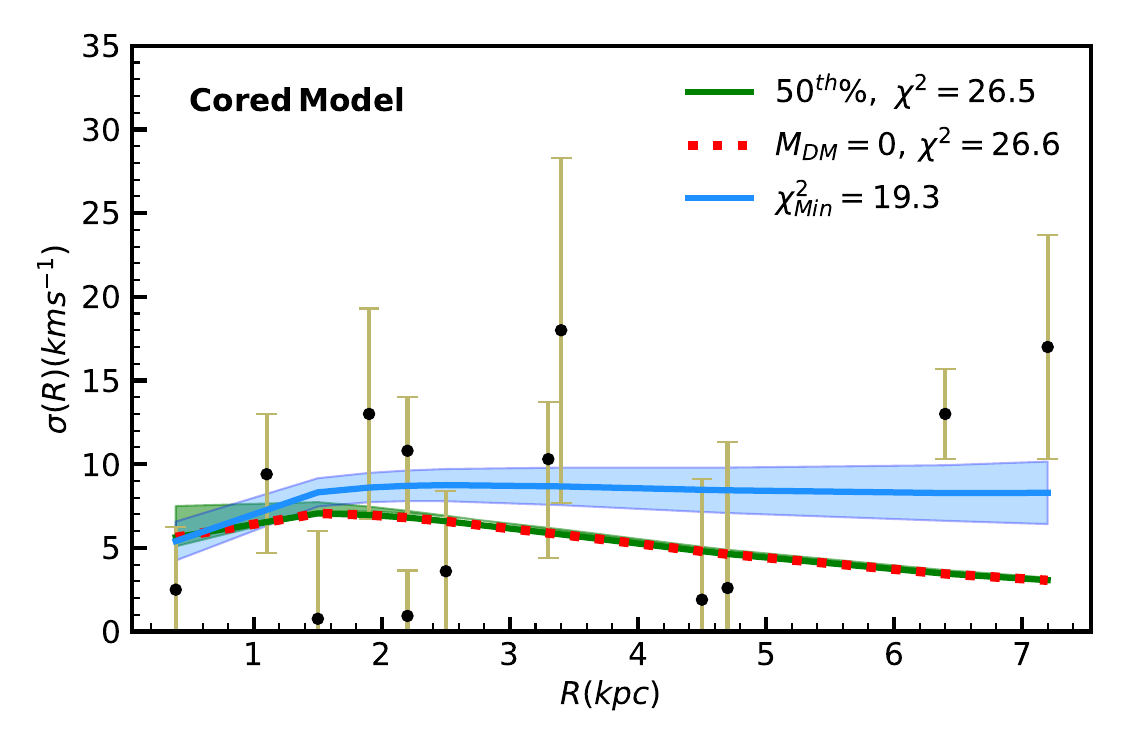}{0.48}{0.}\\                      
\graphicswithlegend{80mm}{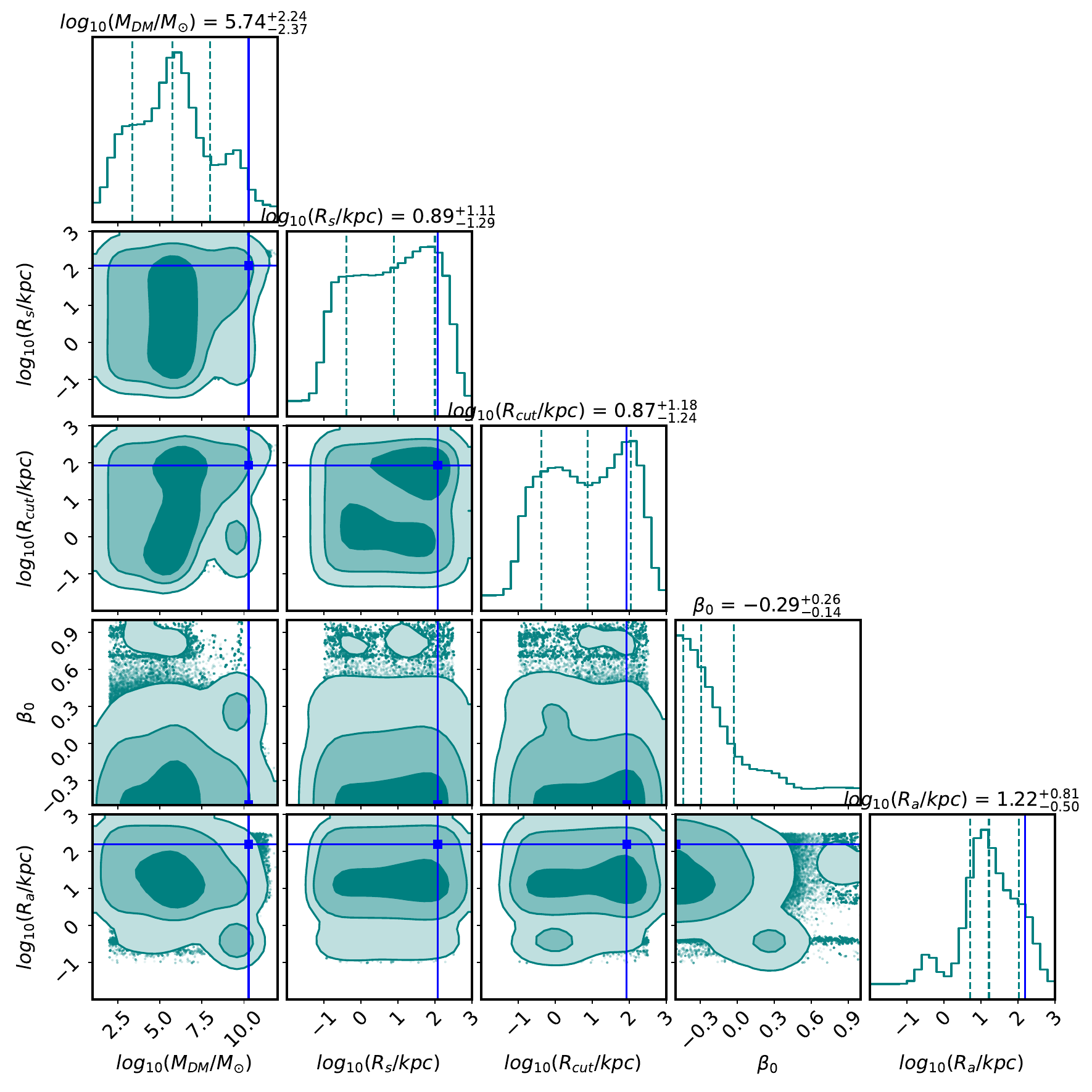}
                       {0.55}{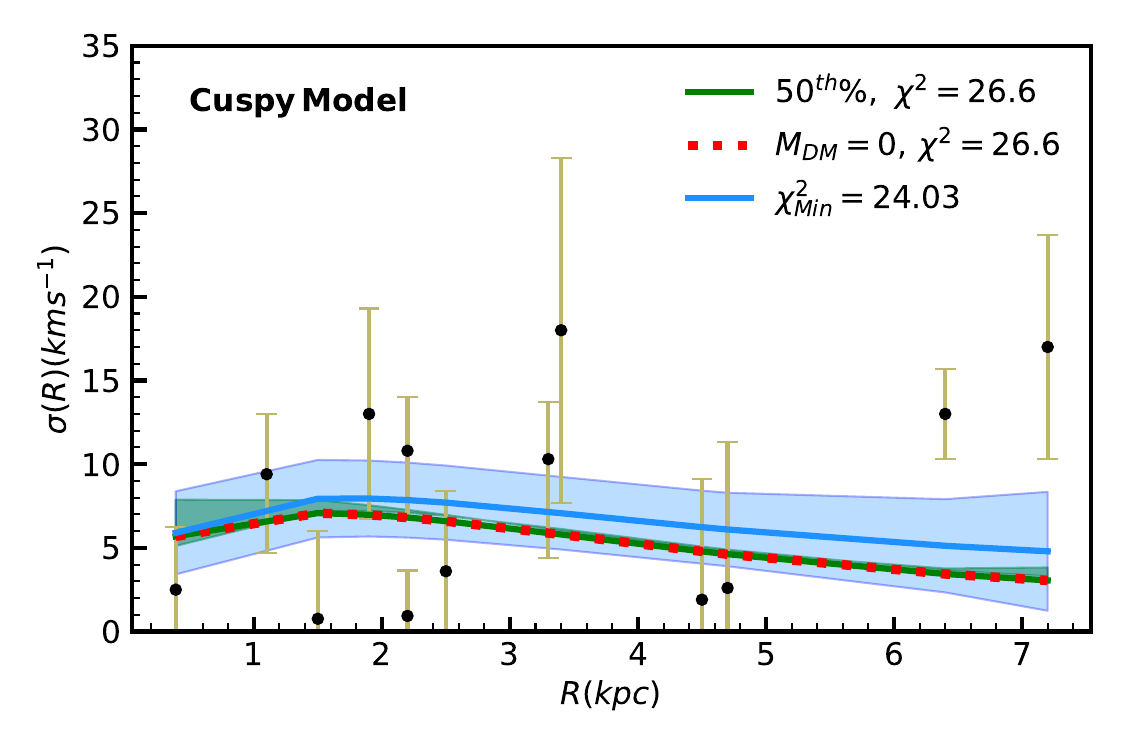}{0.48}{0.}
\caption{Posterior probability distribution corresponding to the cored (top panel) and cuspy (bottom panel) 
dark matter halo. The dashed teal lines depict the $16^{th}$, $50^{th}$, and $84^{th}$ percentiles of the posterior probability distribution.  
{The blue line indicates the parameters corresponding to the model with minimum $\chi^{2}$.}
In the top right corner, the model based on the $50^{th}$ percentile of the posterior and the model with minimum $\chi^{2}$ are 
shown using green and blue lines, respectively. The shaded blue and green regions represent the $1\sigma$ confidence interval. 
The dotted red line depicts the model with zero dark matter ($M_{DM}=0$).}                       
\end{figure}

We computed the posterior probability distribution of the model parameters using the “emcee” Markov Chain
Monte Carlo (MCMC) ensemble sampler \citep{foreman2013emcee}. We ran the sampler with 128 walkers for 30000 iterations 
and rejected the first 2000 iterations to ensure completely mixed chains. We adopted uniform priors on the model parameter:

\begin{enumerate}
\item $2.0$            <  $\log(M_{DM}/M_{\odot})$          <  $12$\\
\item $-1.0$           < $\log(R_{s}/kpc)$                  <  $2.5$\\
\item $-1.0$           <$\log(R_{cut}/kpc)$                 <  $2.5$\\
\item $-0.5$           < $\beta_{0}$                        <  $1.0$\\
\item $-1.0$           < $\log(R_{a}/kpc)$                  <  $2.5$. 
\end{enumerate}
The above priors on the model allowed us to explore the posterior distribution of dark matter halos over a large range of 
dark matter mass $(M_{DM})$ and spatial extent $(R_{s}, \, R_{cut})$. The range of priors also encompasses the stellar 
distributions with tangentially biased anisotropy $(\beta_{0}<0)$, a radially biased 
anisotropy profile $(\beta_{0}>0)$, and isotropic models $(\beta_{0}=0,\, R_{a} \rightarrow \infty)$.
We present the posterior distribution of the model parameters adopting a cored ($\alpha=2, \,\beta=2, \, \gamma=0$) and
cuspy  ($\alpha=1, \, \beta=3, \, \gamma=1$) dark matter halo in the top and bottom panels of Fig. 1, respectively. 
{Upon inspecting the posterior distribution of the cored and cuspy dark matter halo in Fig. 1, 
we find that the mass models are largely unconstrained by the available kinematic data.}\\

The cored dark matter halo a has total dark matter mass, $\log_{10}(M_{DM}/M_{\odot})$,  equal to $7.58^{+2.95}_{-3.43}$ with a 
scale length, $\log_{10}(R_{s}/kpc)$, equal to $1.25^{+0.85}_{-1.32}$ and an outer cutoff, $\log_{10}(R_{cut}/kpc)$, equal to $1.36^{+0.72}_{-1.44}$. 
The stellar distribution is described by a constant tangential anisotropy, with the value of anisotropy at the center $(\beta_{0})$  
equal to $-0.31^{+0.24}_{-0.14}$ and a large anisotropy radius, $\log_{10}(R_{a}/kpc)$, equal to $1.2^{+0.8}_{-0.5}$. The cuspy dark matter model, 
on the other hand, has a total dark matter mass, $\log_{10}(M_{DM}/M_{\odot})$, equal to $5.74^{+2.24}_{-2.37}$ with a scale length, $\log_{10}(R_{s}/kpc)$, equal to $0.89^{+1.11}_{-1.29}$ and an outer cutoff, $\log_{10}(R_{cut}/kpc)$, equal to $0.87^{+1.18}_{-1.24}$. Akin to the cored model, the stellar distribution in the case of the cuspy dark matter halo prefers a constant tangential anisotropy, with the value of 
anisotropy at the center $(\beta_{0})$  equal to $-0.29^{+0.26}_{-0.14}$ and a large anisotropy radius, $\log_{10}(R_{a}/kpc)$, equal to 
$1.22^{+0.81}_{-0.5}$. {At face value, the cored dark matter halo not only has a higher total mass than the cuspy one but also encloses a comparatively higher mass inside $7.6 kpc$; $log_{10}(M_{7.6}/M_{\odot})=6.4$ for the cored halo and 
$log_{10}(M_{7.6}/M_{\odot})=5.6$ for the cuspy halo. However, it is important to emphasize that, within the uncertainties, 
both profiles are consistent with each other.} Based on the posterior distribution, we find that the effective 
contribution to the dynamical mass inside $7.6kpc$ is due to the stellar distribution. Further, we also note that both the cored and 
cuspy dark matter models have a large spatial extent; in other words, a large value of $R_{s}$ and $R_{cut}$. 
Similar to $R_{s}$ and $R_{cut}$, the anisotropy radius, $R_{a}$, also extends beyond the spatial extent of the 
stellar distribution, indicating that $\beta(r)$ is constant equal to $\beta_{0}$ within the galaxy. {Further, it is evident from 
the posterior distribution that the available data does not allow us to obtain stringent constraints on the halo properties; both models also have a comparable $\chi^{2}$ value.} {Moreover, the $\chi^{2}$ for the model based on the median value of the 
posterior probability distribution is comparable to the models with $M_{DM}=0$, given by the dotted red line in Fig. 1}. This indicates 
that NGC 1052 - DF2 has a sufficient stellar density to reproduce the observed
dispersion in the inner region of the galaxy at the same levels as models with $M_{DM}\leq10^{8}$, without having to invoke the contribution of dark matter to the total potential. \cite{2022MNRAS.512.3230M}, 
in their study of the gas-rich UDG AGC 114905, find that dark matter halos that follow 
the standard concentration-mass relation fail to reproduce the observed rotation curve of the galaxy. 
\cite{2022MNRAS.512.3230M} find that only a dark matter halo with an extremely low concentration parameter agrees 
with the data (see also \cite{mancera2020robust, 2021ApJ...909...20S}). This indicates that it is hardly possible to 
constrain the shape of the dark matter halo $(core/cuspy)$ or the concentration parameter for arbitrary small values of 
dark matter mass.

While we cannot resolve the degeneracy between the cored and cuspy dark matter model, 
we note that the posterior probability distribution for a cored halo tends to favor a higher total dark matter mass than the cuspy dark matter halo. 
{Our analysis suggests that mass models of NGC 1052 - DF2 with a cored dark matter model can accommodate a higher dark matter within 
the spatial extent of the galaxy defined by the availability of the kinematic data (7.6 kpc).} Using cosmological zoom-in simulations, 
\cite{di2014dependence} show that the core formation efficiency depends on the stellar-to-halo mass ratio and not 
just the halo mass. \cite{di2014dependence} point out that higher SHMR feedback can drive the expansion of the dark matter halo 
and generate the cored profiles. Cored dark matter profiles are not uncommon in low-mass systems. \cite{read2016dark} using high-resolution 
simulations show that a dwarf galaxy with a stellar mass equal to $10^{6}M_{\odot}$ and a dark matter mass equal to $10^{9}M_{\odot}$ can 
form dark matter cores comparable to the stellar half mass radius.

We find that the posterior probability distribution for our model is non-Gaussian, bimodal, and highly complex; thus, unlike the usual case, the $50^{th}$ percentile does not correspond to the minimum $\chi^{2}$ solution. In the top right corner of Fig. 1, 
we show the model with minimum $\chi^{2}$ for the cored halo dark matter halo model (top panel) and the cuspy one
(bottom panel) using the blue line. The cored dark matter halo model that minimizes the $\chi^{2}$ has $\chi^{2}=19.3$. The cored 
$\chi^{2}_{Min}$ model has a total dark matter mass, $\log_{10}(M_{DM}/M_{\odot})$, 
equal to $10.48^{+0.4}_{-1.6}$ with a scale length, $\log_{10}(R_{s}/kpc)$, equal to $1.3^{+1.0}_{-0.12}$ and an outer 
cutoff, $\log_{10}(R_{cut}/kpc)$, equal to $1.42^{+0.97}_{-0.12}$. The stellar distribution is described by a constant tangential anisotropy, 
with the value of anisotropy at the center $(\beta_{0})$  equal to $-0.49^{+0.46}_{-0.01}$ and an  anisotropy radius, 
$\log_{10}(R_{a}/kpc)$, equal to $1.64^{+0.35}_{-0.64}$. On the other hand, the cuspy dark matter model that minimizes the 
$\chi^{2}$ has $\chi^{2}=24$. The cuspy $\chi^{2}_{Min}$ model has a total dark matter mass, $\log_{10}(M_{DM}/M_{\odot})$, equal to $10.27^{+0.37}_{-1.8}$ with a scale length, $\log_{10}(R_{s}/kpc)$, equal to $2.07^{+0.42}_{-0.69}$ and an outer cutoff, $\log_{10}(R_{cut}/kpc)$, equal to $1.93^{+0.56}_{-0.43}$. Akin to other models, the $\chi^{2}_{Min}$ model also has a constant tangential anisotropy, with the value of anisotropy at the center $(\beta_{0})$  equal to $-0.49^{+0.55}_{-0.01}$ and an anisotropy radius, $\log_{10}(R_{a}/kpc)$, equal to $2.19^{+0.29} _{-1.59}$. The values of $R_{s}$ and $R_{cut}$ are unrealistically large for the cuspy dark matter halo, almost comparable to the projected distance between NGC 1052 -DF2 and the luminous elliptical galaxy NGC 1052 \citep{van2018galaxy}. The cored dark matter halo encloses dark matter mass, $log_{10}(M_{7.6}/M_{\odot})=9.1$, inside 7.6kpc, whereas the cuspy model encloses $log_{10}(M_{7.6}/M_{\odot})=8.5$. 
The $\chi^{2}_{Min}$ models not only have a total dark matter mass comparable to values predicted by the SHMR but also enclose a higher dark matter mass inside the spatial extent of the galaxy ($\sim 7.6kpc$) up to which the kinematic and photometric data are available. Further, we note that the model based on a cored dark matter halo consistently explains the small velocity dispersion observed in 
the inner region and the larger velocity dispersion in the outer region of the galaxy. 
{The models that minimize the $\chi^{2}$ suggest that cored dark matter models can support higher dark matter masses consistent with the SMHR. However, the posterior distribution of the parameters is biased toward data points in the inner region with a low velocity dispersion. The data points in the inner region can be accounted for by mass models with $M_{DM}=0$ (see the dotted red line in Fig. 1).
It is crucial for the mass models to explain the velocity dispersion in the outer region because the velocity dispersion in the outer region traces the total potential of the galaxy. Accurate estimates of dark matter mass will only be achieved when the models consistently explain the velocity dispersion in both the inner and outer regions of the galaxy, as is highlighted by the minimum $\chi^{2}$ model. Therefore, we note that the available kinematic data is insufficient to derive stringent constraints on the dark matter models.} 

\section{Discussion}

\subsection{Degeneracy in model parameters}

The posterior probability distribution for our models is highly complex and deviates from Gaussianity. Further, we note that the posterior distribution is highly bimodal and does not show a clear maximum in the 
parameter space. Thus, the model parameters that minimize the $\chi^{2}$ do not correspond to the median value of 
the posterior probability distribution. The cored dark matter models that minimize the $\chi^{2}$ align with SHMR, whereas 
the models based on the median values of the posterior probability distribution minimize the contribution of dark matter within the 
observed spatial extent of the galaxy. This suggests that the current data is insufficient for deriving strong constraints on the dynamical 
models of NGC 1052 - DF2. Given the limited data, we draw the following observation regarding the inherent degeneracies within the model based on the posterior distribution of models obtained through MCMC analysis, as well as the models with the minimum chi-squared ($\chi^{2}$) value:

\begin{enumerate}
    \item Models with tangential anisotropy describe the observed stellar dispersion profile better than the isotropic and radially biased models.
    
    \item Mass models have a large anisotropy radius, indicating that models with constant anisotropy  $(\beta(r)=\beta_{0})$ 
    describe the observed dispersion profile better than the models with a radially varying anisotropy parameter.

    \item Mass models with a large scale length and a large outer cutoff radius explain the observed stellar dispersion profile better 
    than those with a smaller scale length and outer cutoff.

    \item  The cored dark matter models 
    can accommodate a higher total dark matter mass consistent with the SHMR and, at the same time, explain the small stellar velocity dispersion at the center and larger velocity 
    dispersion at the outer radius. Also, studies by \cite{ogiya2018tidal} have shown in their 
    N-body simulations that a cored dark matter halo explains the observed properties of the NGC 1052-DF2 better than the cuspy 
    dark matter halo.
    
    \item Mass models with a cuspy dark matter halo with $M_{DM} >10^{8} M_{\odot}$ require an unrealistically large scale length and outer cutoff, comparable to 
    the projected distance between NGC 1052 - DF2 and the luminous elliptical galaxy NGC 1052 - DF2. Further, the cuspy halo models based on a large 
    scale length and outer cutoff are ineffective at explaining the observed velocity dispersion in the outer region.
    Thus, we can rule out the possibility of a cuspy dark matter halo in mass models of $M_{DM} >10^{8} M_{\odot}$.

    \item The significance of dark matter within the observed spatial extent of a galaxy becomes apparent only when mass models 
    successfully account for the stellar dispersion in the outer region because the velocity dispersion in the outer region traces 
    the total potential of the galaxy. The velocity dispersion in the inner regions can be explained by models with zero dark matter that have $\chi^{2}$ comparable to the various dark matter models. Thus, any mass model of NGC 1052 - DF2 should consistently 
    explain the velocity dispersion in both inner and outer regions.\\
\end{enumerate}

\subsection{Whether SHMR-based models can explain the observed stellar dispersion}
Our analysis indicates that the cored dark matter halo, with a dark matter mass of $\log_{10}(M_{DM}/M_{\odot}) = 10.48^{+0.4}_{-1.6}$, 
exhibits a smaller $\chi^{2}$ equal to 19.3 compared to the models assuming zero dark matter and the models based on the median value of the posterior probability distribution. The latter models have a comparable $\chi^{2} \sim 26$. Further, the cored model that yields a minimum 
$\chi^{2}$ also encloses a dark matter mass, $log_{10}(M_{7.6}/M_{\odot})=9.1$, inside 7.6kpc, maximizing the contribution of the dark 
matter to the total potential. {The results suggest that we cannot rule out the possibility that NGC 1052-DF2 might be a typical dark matter-dominated system, consistent with galaxy formation and evolution models.} {Our findings suggest that NGC 1052-DF2 within the uncertainties 
could align with a higher dark matter mass, as was predicted by the SHMR, provided that the dark matter profile has a large scale length and a large cutoff radius.}
Our findings indicate that NGC 1052 - DF2 not only 
hosts an ultra-diffuse stellar distribution but possibly hosts an ultra-diffuse dark matter distribution as well.
We note that adopting an SHMR-based dark matter model for NGC 1052-DF2 would require an explanation for the 
large scale length and outer cutoff radius, which is limited by the physical extent of the galaxy itself. 
Usually, in rotation-supported systems, strong constraints on the shape of dark matter halo can be obtained 
using the stellar distribution in the inner region and the distribution of neutral hydrogen in the outer region (for example, see  \citep{2001ApJ...552L..23D, 2017MNRAS.467.2019R, 2017MNRAS.466.4159I, 2022MNRAS.514.3329M}). 
However, the paucity of neutral hydrogen \citep{chowdhury2018dearth,sardone2019constraints} in this galaxy and 
the limited radial extent of the optical tracers \citep{montes2021disk} makes it challenging to explain the 
large scale length and outer cutoff radius needed for an SHMR-based model. A straightforward way of testing the SHMR-based models would be to study the properties of dark matter halos in galaxy formation simulations. 
Recent galaxy formation simulations like Auriga \citep{liao2019ultra}, NIHAO \citep{di2017nihao}, and Romulus 
\citep{van2022s} study the formation of UDGs in the cosmological scenario. Thus, studying 
the structure and extent of the dark matter halos of the UDGs in these simulations may provide clues 
for understanding if SHMR-based dynamical models for UDGs warrant a large scale length and a large 
outer cutoff radius. In their ultra-deep imaging study of NGC 1052-DF2, \cite{montes2021disk} 
find that stellar distribution in NGC 1052- DF2 remains unperturbed up to a considerable distance and lacks the signature 
of tidal disturbances like tidal tails. Using N-body simulations, \cite{ogiya2018tidal} suggest that the stars are 
tightly bound to the central dark matter core and are thus less susceptible to the tidal force of the massive host galaxies, whereas the 
dark matter on the outskirts is more loosely bound and is prone to tidal stripping. 
Thus, tidal stripping may be a possible mechanism that explains the large scale lengths and 
the outer cutoff radius that we find in SHMR-based models.\\

\subsection{Effect of distance on dark matter mass estimates}
The dark matter mass in NGC 1052 - DF2 is distance-dependent, since the distance
to the galaxy is required to derive the stellar mass from the 2D stellar distribution. 
Studies by \cite{van2018distance} estimate the distance to NGC 1052 - DF2 
as being equal to $19 \pm 1.7 Mpc$, whereas another independent study by 
\cite{trujillo2019distance} indicates a much shorter distance equal to $13Mpc$. Using a distance equal 
to $13Mpc$, \cite{trujillo2019distance} find that NGC 1052 - DF2 is akin to an ordinary 
low-surface-brightness galaxy with a stellar mass equal to $6\times 10^{7}M_{\odot}$ and 
an effective radius equal to $1.4kpc$. The smaller stellar mass brings NGC 1052 - DF2 closer 
to the SHMR, allowing for a much larger dark matter contribution to the total mass. \cite{trujillo2019distance} 
show that the cored and cuspy dark matter halo models, with total masses equal to $10^{9}M_{\odot}$ and $10^{9.6}M_{\odot}$, 
respectively, result in a dynamical mass equal to $4.3\times10^{8}M_{\odot}$ within $5kpc$, making NGC 1052 - DF2 a dark-matter-dominated galaxy. 
Interestingly, \cite{trujillo2019distance}, akin to this study, find that a cored dark matter model 
with a large scale length can fit the observed dispersion with a higher dark matter mass. A 
smaller distance and a smaller stellar mass provide a natural solution to the dark matter conundrum; a smaller stellar mass naturally allows more room for dark matter. To gauge the effect of the smaller distance equal to 13 Mpc 
on the dynamical models, we derived the dynamical models assuming the photometric parameters derived by \cite{trujillo2019distance}. 
The smaller distance effectively changes the relative distance between the kinematic tracers and the spatial extent of the galaxy. At a distance equal to 20 Mpc, the kinematic tracers extend up to 7.6 kpc, whereas in the case of 13 Mpc, the spatial extent of kinematic tracers is reduced to 4.7 kpc. 
{We show the dynamical models corresponding to minimum $\chi^{2}$ assuming a distance equal to 13 Mpc in Fig. 2 and show the 
posterior distribution in Fig. 3 in the appendix. Similar to the mass models at 20 Mpc, the posterior for the mass 
models at 13 Mpc is also biased toward the kinematic data in the inner region, which can be explained without the contribution of dark matter (see the dotted red line in Fig. 2). } 
The cored dark matter halo, shown using a solid green line, has a total dark matter mass, $\log_{10}(M_{DM}/M_{\odot})$, equal to $10.32^{+0.54}_{-1.88}$ with a scale length, $\log_{10}(R_{s}/kpc)$, equal to $0.86^{+0.89}_{-0.29}$ and an outer cutoff, $\log_{10}(R_{cut}/kpc)$, equal to $1.51^{+0.89}_{-0.31}$. The stellar distribution is described by a constant tangential anisotropy, with the value of anisotropy at the center $(\beta_{0})$  
equal to $-0.49^{+0.46}_{-0.01}$ and a large anisotropy radius, $\log_{10}(R_{a}/kpc)$, equal to $0.74^{+1.16}_{-0.04}$. The cored dark 
matter encloses $log_{10}(M_{4.7}/M_{\odot})=8.7$ inside 4.7 kpc, indicating a significant contribution of the dark matter to 
the total mass budget inside 4.7kpc. The cored dark matter model has $\chi^{2}$ equal to 16.9. The cuspy dark matter model is shown using a solid blue line and has a total dark matter mass, $\log_{10}(M_{DM}/M_{\odot})$, equal to $9.38^{+0.54}_{-1.99}$ with a scale length, $\log_{10}(R_{s}/kpc)$, 
equal to $0.88^{+0.78}_{-0.3}$ and an outer cutoff, $\log_{10}(R_{cut}/kpc)$, equal to $2.16^{+0.33}_{-0.66}$. The cuspy dark matter halo also 
prefers a constant tangential anisotropy, with the value of anisotropy at the center $(\beta_{0})$ equal to $-0.49^{+0.52}_{-0.01}$ and an anisotropy 
radius, $\log_{10}(R_{a}/kpc)$, equal to $0.98^{+0.62}_{-0.38}$. The cuspy dark matter encloses $log_{10}(M_{4.7}/M_{\odot})=8.1$ inside 4.7 kpc, and has $\chi^{2}$ equal to 23.6. Similar to our analysis at a distance equal to 20 Mpc, we find that even at 13 Mpc, the cored dark matter model explains the observed velocity dispersion better than the cuspy one. The cored dark matter, besides having smaller $\chi^{2}$, consistently explains the observed dispersion profile in both the inner and outer regions. The cuspy dark matter halo model and the model with zero dark matter shown using a dotted red line explain 
the small velocity dispersion observed in the inner region but does not explain the relatively larger velocity dispersion in the outer region, which effectively traces the total dynamical mass of the system. {The models at 13 Mpc and 20 Mpc exhibit qualitatively similar behavior. Our analysis of mass models adopting a distance equal to 13 Mpc shows that the posterior is biased toward kinematic data in the inner region. The kinematic data in the inner region can be explained without invoking the contribution of 
the dark matter (see the dotted red line in Fig. 1 and Fig. 2). But the significance of dark matter within the observed spatial extent of a galaxy becomes apparent only when mass models successfully account for the stellar dispersion in the outer region because the velocity dispersion in the outer region traces the 
total potential of the galaxy. Thus, similar to our findings for 20 Mpc, we emphasize that the kinematic data is insufficient to derive stringent constraints on mass models.}

\begin{figure}
\resizebox{95mm}{65mm}{\includegraphics{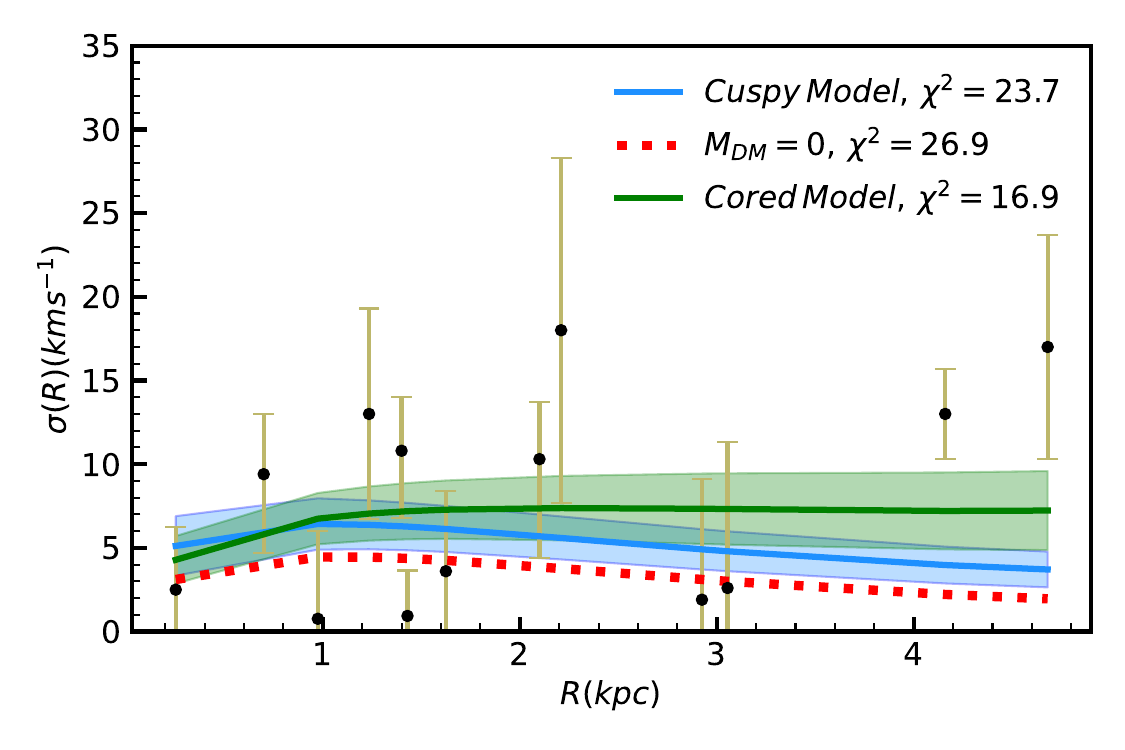}} 
\caption{Dynamical models of NGC 1052 -DF2, assuming a distance equal to 13Mpc. The solid green and blue lines represent the 
models derived using the cored and cuspy dark matter halos, respectively. The shaded green and blue regions depict a $1\sigma$ confidence interval.
The dotted red line indicates a model with zero dark matter.}
\end{figure}
\section{Conclusions}
In this work, we have studied the dynamical model of UDG NGC 1052 - DF2 based on the anisotropic distribution function using 
the photometric and kinematic data available in the literature. NGC 1052 - DF2 challenges our current understanding of galaxy formation models in the $\Lambda CDM$ scenario. Previous studies have shown that NGC 1052 - DF2 is deficient in dark matter and is in conflict with mass 
models based on the standard SHMR. 
We find that the stellar distribution is tangentially biased, and that the anisotropy parameter is 
constant with radius. Further, we find that models with a dark matter mass  $<10^{8}M_{\odot}$ are riddled with degeneracy between the cored and cuspy halo, 
but models with a higher dark matter mass that explain the observed kinematics
prefer cored halos with a large scale length and a large outer cutoff radius. The dark matter models consistent with SHMR need an extraordinarily large scale length (20kpc) and a large outer cutoff radius (26kpc), much larger than the 
spatial extent at which kinematic and photometric data are available. This provides 
a unique opportunity to study the properties of dark matter halos of UDGs in galaxy formation simulations like 
AURIGA, NIHAO, and ROMULUS and understand if UDGs are characterized by a large scale length and outer cutoff radius. 
{Our results suggest that NGC 1052 - DF2 might not only have an ultra-diffuse stellar distribution but might potentially host an ultra-diffuse dark matter distribution compatible with the standard SHMR, as has been predicted by galaxy formation and evolution models, within uncertainties on the available kinematic data. Our results emphasize that the current kinematic data are insufficient for obtaining precise constraints on the  dynamical models of NGC 1052 - DF2.}

\begin{acknowledgements}
Aditya would like to thank the referee for their insightful comments that improved the quality of this manuscript.
\end{acknowledgements}

\small{\bibliographystyle{aa}}
\begin{itemize}
    \item \bibliography{example}.
\end{itemize}

\newpage
\begin{appendix}
\section{Posterior distribution of models at a distance equal to 13 Mpc}
In the appendix, we have shown the posterior distribution corresponding to the cored and cuspy dark matter halo, adopting 
a distance to NGC 1052 - DF2 equal to 13 Mpc. 
\begin{figure}
\hspace{-3mm}
\graphicswithlegend{80mm}{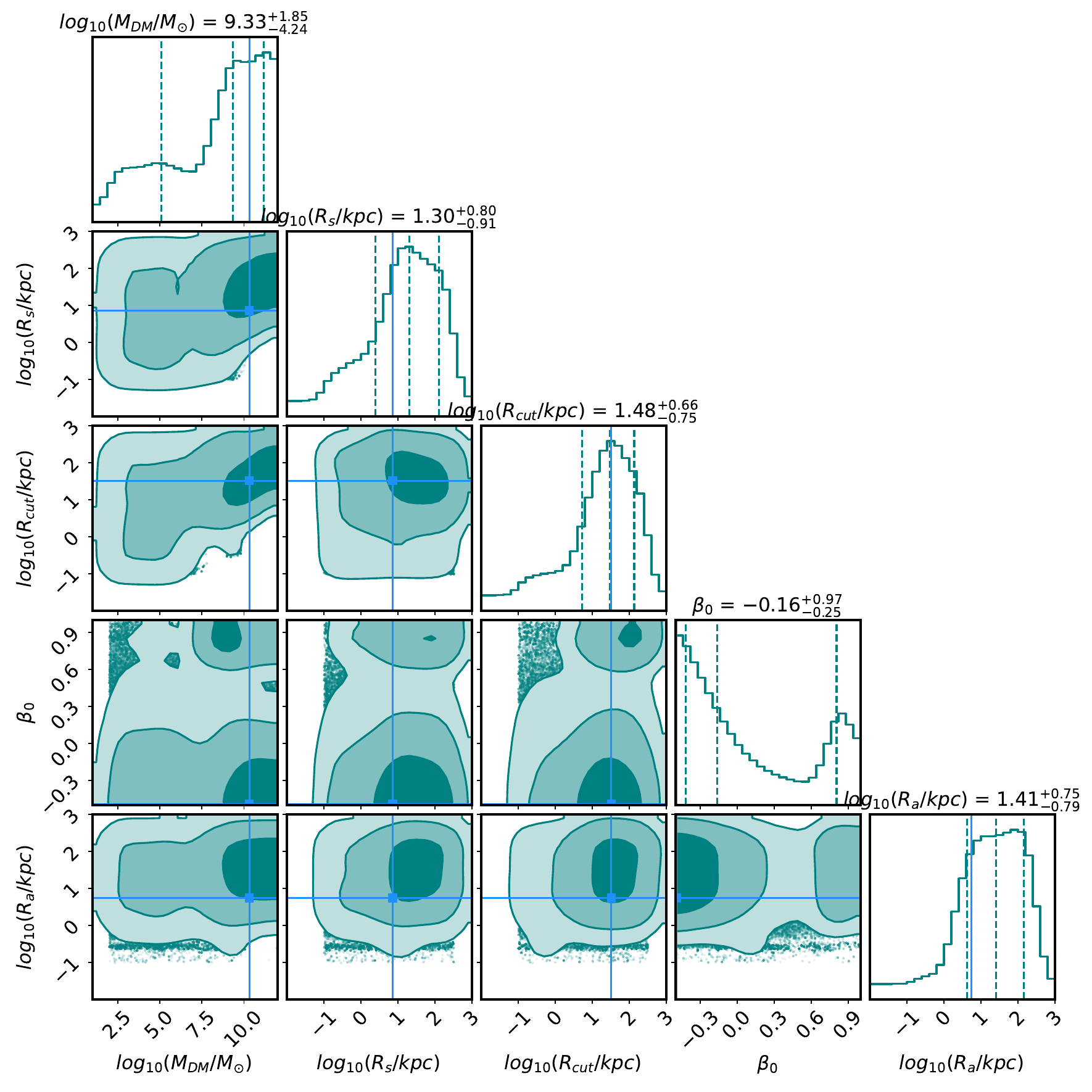}
                       {0.55}{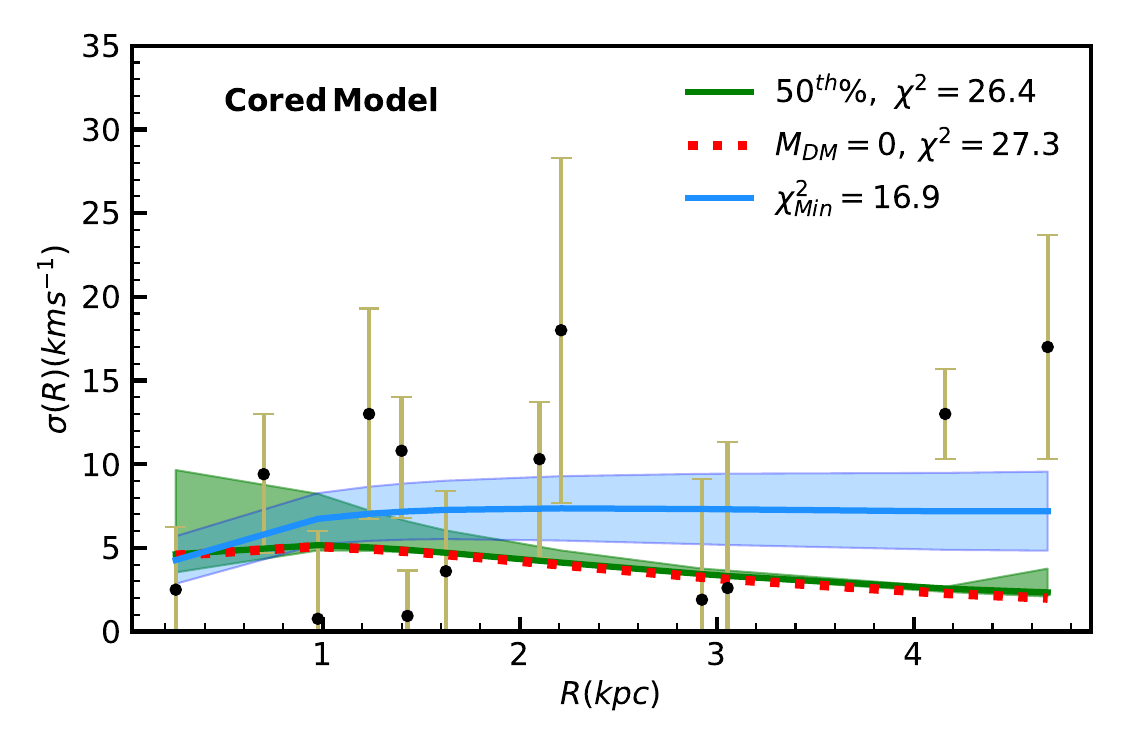}{0.48}{0.}\\                      
\graphicswithlegend{80mm}{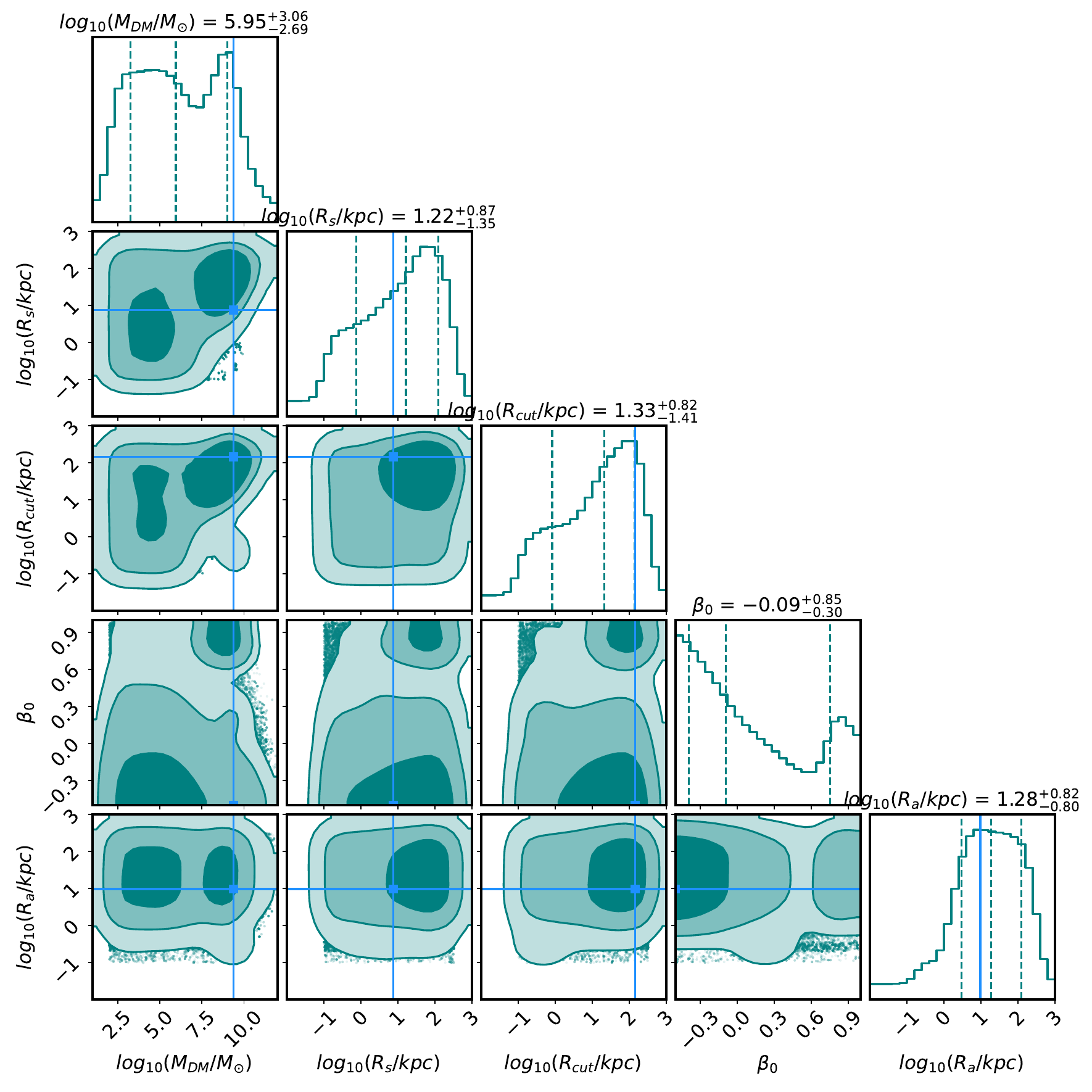}
                       {0.55}{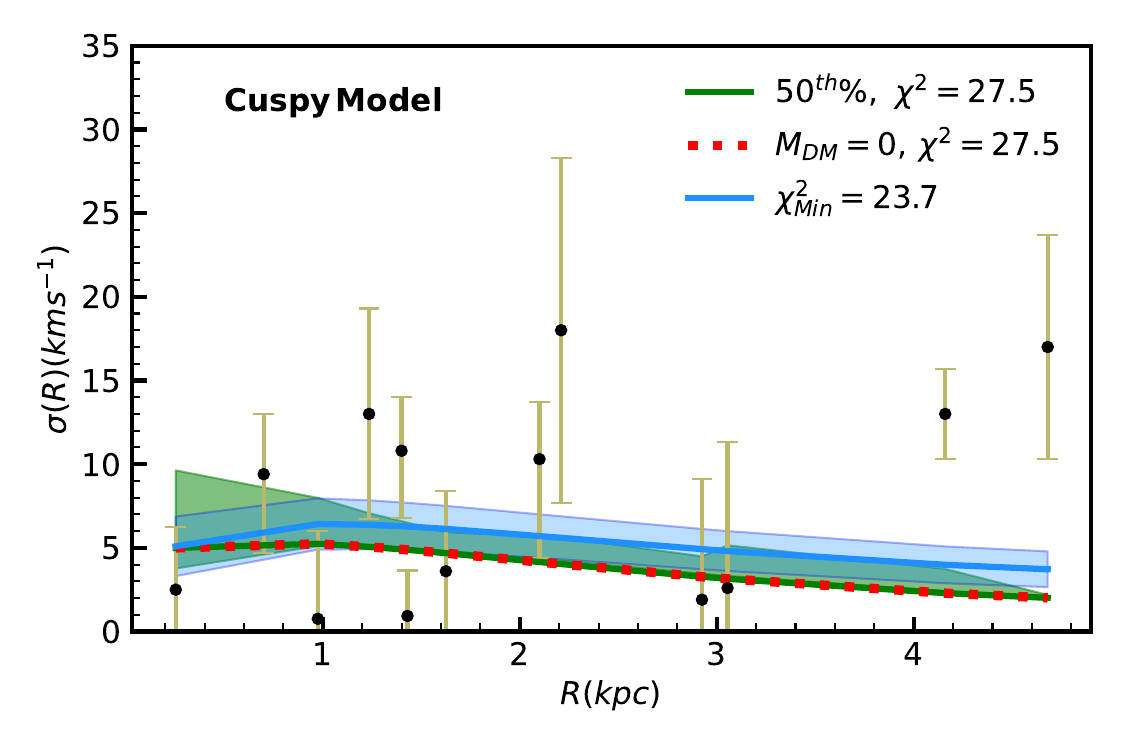}{0.48}{0.}
\caption{{Posterior probability distribution corresponding to the cored (top panel) and cuspy (bottom panel) 
dark matter halo, adopting a distance to NGC 1052 - DF2 equal to 13 Mpc. The dashed teal lines depict the $16^{th}$, $50^{th}$, and $84^{th}$ percentiles of the posterior probability distribution.  
The blue line indicates the parameters corresponding to the model with minimum $\chi^{2}$.
In the top right corner, the model based on the $50^{th}$ percentile of the posterior and the model with minimum $\chi^{2}$ are 
shown using green and blue lines, respectively. The shaded blue and green regions represent the $1\sigma$ confidence interval. 
The dotted red line depicts a model with zero dark matter ($M_{DM}=0$).}}                       
\end{figure}
\end{appendix}

\end{document}